\documentclass[epj]{webofc}
\usepackage[varg]{txfonts}   
\wocname{EPJ Web of Conferences}
\woctitle{CONF12}
\usepackage{color}

\usepackage[normalem]{ulem}  

\newcommand{\be}{\begin{equation}}
\newcommand{\ee}{\end{equation}}
\newcommand{\bea}{\begin{eqnarray}}
\newcommand{\eea}{\end{eqnarray}}

\begin{document}
\selectlanguage{english}
\title{Phases of dense matter with holographic instantons}
%
%

\author{Florian Preis\inst{1}\fnsep\thanks{\email{fpreis@hep.itp.tuwien.ac.at}} \and
        Andreas Schmitt\inst{2}\fnsep\thanks{\email{a.schmitt@soton.ac.uk}}
}

\institute{Institut f\"{u}r Theoretische Physik, Technische Universit\"{a}t Wien, 1040 Vienna, Austria
\and
          Mathematical Sciences and STAG Research Centre, University of Southampton, Southampton SO17 1BJ, United Kingdom
}

\abstract{%
  We discuss nuclear matter and the transition to quark matter in the decompactified limit of the Sakai-Sugimoto model. Nuclear matter is included through 
   instantons on the flavor branes of the model. Our approximation is based on the flat-space solution, but we allow for a dynamical instanton width and deformation and compute the energetically preferred 
  number of instanton layers in the bulk as a function of the baryon chemical potential. We determine the regions in parameter space where the binding energy of nuclear matter is like in QCD, and compute 
  the phase diagram in the plane of temperature and chemical potential. }
\maketitle
\section{Introduction}
\label{intro}

The Sakai-Sugimoto model \cite{Sakai:2004cn,Sakai:2005yt} is a top-down realization of the gauge/string duality \cite{Maldacena:1997re}, which is, in a certain limit, dual to large-$N_c$ QCD. 
Although only the opposite limit is accessible, that of large 't Hooft coupling $\lambda$, where the strict duality to QCD is lost, it has been successfully applied to meson, baryon, and 
glueball properties in the vacuum \cite{Sakai:2004cn,Sakai:2005yt,Hata:2007mb,Imoto:2010ef,Brunner:2015yha}. The results suggest that the model can capture certain non-perturbative features that are very hard, if not impossible, to reproduce, unless the full theory is evaluated, for instance on the lattice. 
Therefore, the model constitutes one of the most promising applications of the gauge/string duality in the context of QCD, although it always has to be kept in mind that we usually work in the limit of infinite coupling and  
infinite number of colors, and that currently, at best, we can compute small corrections away from that limit. 

A natural question is whether the Sakai-Sugimoto model can also be applied to medium physics, i.e., to QCD at finite temperature and density. Hot QCD, at zero or small baryon chemical potential, as encountered in relativistic 
heavy-ion collisions, has been approached extensively with holographic methods \cite{casalderrey2014gauge}. Mostly, for this purpose, the original and best established version of the gauge/string correspondence has been employed, with ${\cal N}=4$ supersymmetric Yang-Mills theory being the dual field theory. Even though this theory is different from QCD, its application to the physics of heavy-ion collisions, or, more precisely, 
the comparison of its predictions with the experimental results of the real-world quark gluon plasma, has turned out to be fruitful. 

How about the complementary region, that of low temperature and large chemical potential?
This form of matter is interesting from a phenomenological point of view as well, because cold and dense matter exists inside compact stars, be it as nuclear matter 
(neutrons and protons in the simplest form, probably including Cooper pairing) or as deconfined quark matter (three-flavor quark matter, also likely to form Cooper pairs) \cite{Schmitt:2010pn}.  
Strongly coupled dense matter presents
an enormous theoretical challenge: it is currently out of reach for lattice QCD -- although there is progress in that direction \cite{Aarts:2015tyj}, 
and its phase structure is most likely to be very rich, not unlike ordinary condensed matter systems \cite{Alford:2007xm}. Moreover, 
since compact stars have a density profile, they probe a wide region of the QCD phase diagram, possibly including both confined and deconfined matter. Even within conventional field-theoretical models that are already 
simplified tremendously compared to full QCD, it is difficult to incorporate both nuclear and quark matter, and current efforts to compute the equation of state of ultra-dense matter often are restricted to 
one form of matter or have to rely on a patchwork of different approaches, which is unsatisfying from a theoretical point of view. 
Holographic approaches exist (although by far not as numerous as in the heavy-ion context), but they usually 
rely on rudimentary descriptions of nuclear matter and do not contain quark matter at all \cite{Ghoroku:2013gja,Kim:2014pva}, 
or they use conventional, field-theoretical descriptions of nuclear matter and combine them with a holographic approach for (supersymmetric) quark matter \cite{Hoyos:2016zke}.  
The Sakai-Sugimoto can potentially overcome some of these problems. Although, as argued above, we do not expect the model in its accessible limit to reflect real-world QCD in any rigorous way, it {\it does}
know about a chiral and deconfinement phase transition, about nucleons and quark matter at strong coupling. Therefore, it is worthwhile improving existing approximations, eventually constructing a 
holographic equation of state which may be useful as an approach to dense matter complementary to more traditional methods. 

\begin{figure}[t]
\centering
\includegraphics[width=.9\textwidth]{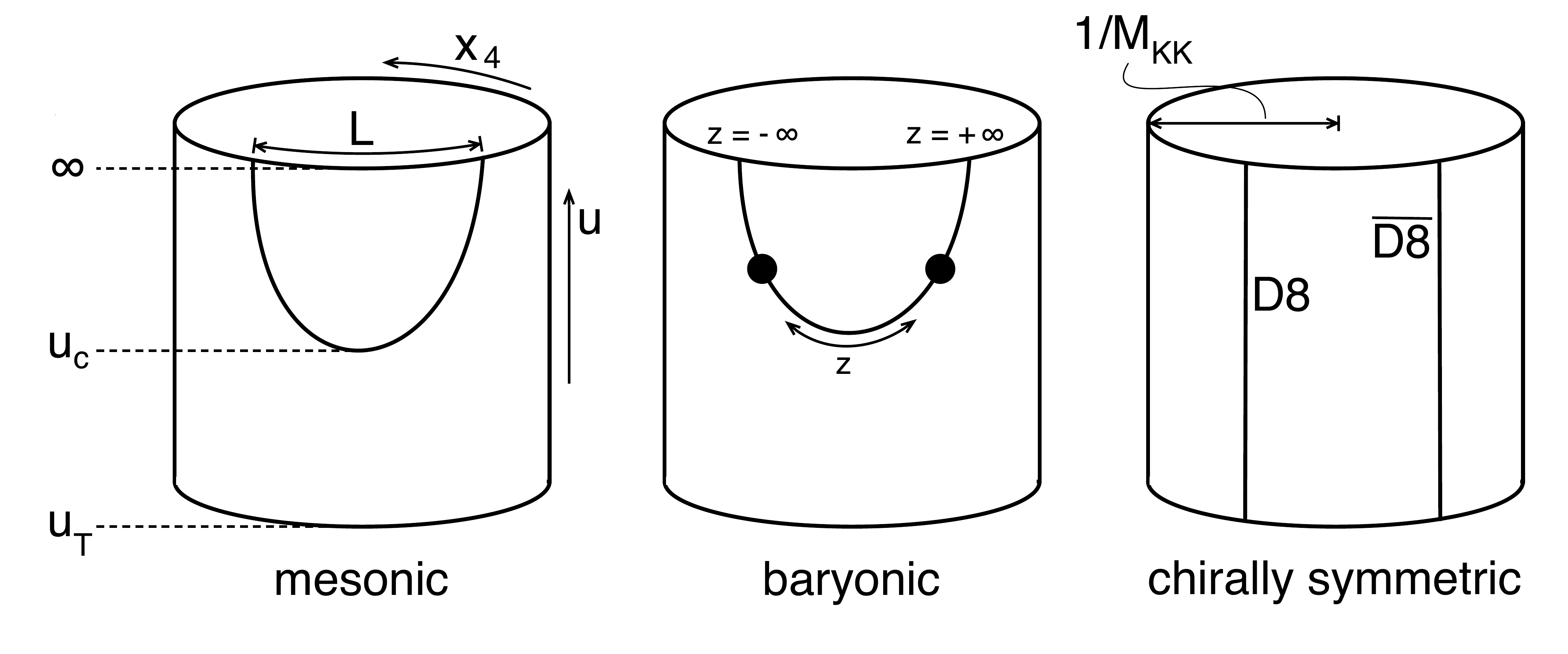}
\caption{Chirally broken phases with and without baryons, and chirally symmetric phase in the deconfined geometry of the Sakai-Sugimoto model, shown in the subspace spanned by the holographic direction $u\in[u_T,\infty]$,
where $u_T$ is given by the temperature (on the connected flavor branes this direction is, alternatively, parameterized by $z\in[-\infty,\infty]$), 
and the compactified extra dimension $x_4$ with radius $M_{\rm KK}^{-1}$.
The black circles in the baryonic phase represent two instanton layers. In the pointlike approximation, with a single instanton layer located at the tip
of the connected branes $u_c$, the profile of the branes acquires a cusp at that point (hence the notation $u_c$ which is used here, even though the embedding is smooth in the case of finite-size instantons).
The calculations in the present work are performed in the decompactified limit, $L\ll \pi/M_{\rm KK}$. 
}
\label{fig:3cylinders}    
\end{figure}

Our calculation is done in the so-called decompactified limit of the model. 
In this limit, the asymptotic separation of the D8- and $\overline{\rm D8}$-branes $L$ is small compared to the radius of the compactified extra dimension, for a schematic picture of the geometry of the model see  
Fig.\ \ref{fig:3cylinders}. Moving the flavor branes asymptotically from their maximal separation (this is the "original" version of the model) very close together can be thought of as 
going from large-$N_c$ QCD to a different dual theory 
in which the gluon dynamics decouples, not unlike a Nambu--Jona-Lasinio model \cite{Antonyan:2006vw,Davis:2007ka,Preis:2012fh}. The reason for working in this limit is that now the chiral phase transition depends on the 
baryon chemical potential, even without computing corrections away from the infinite-$N_c$ limit. 

\begin{figure}[t]
\centering
\hbox{\includegraphics[width=.5\textwidth]{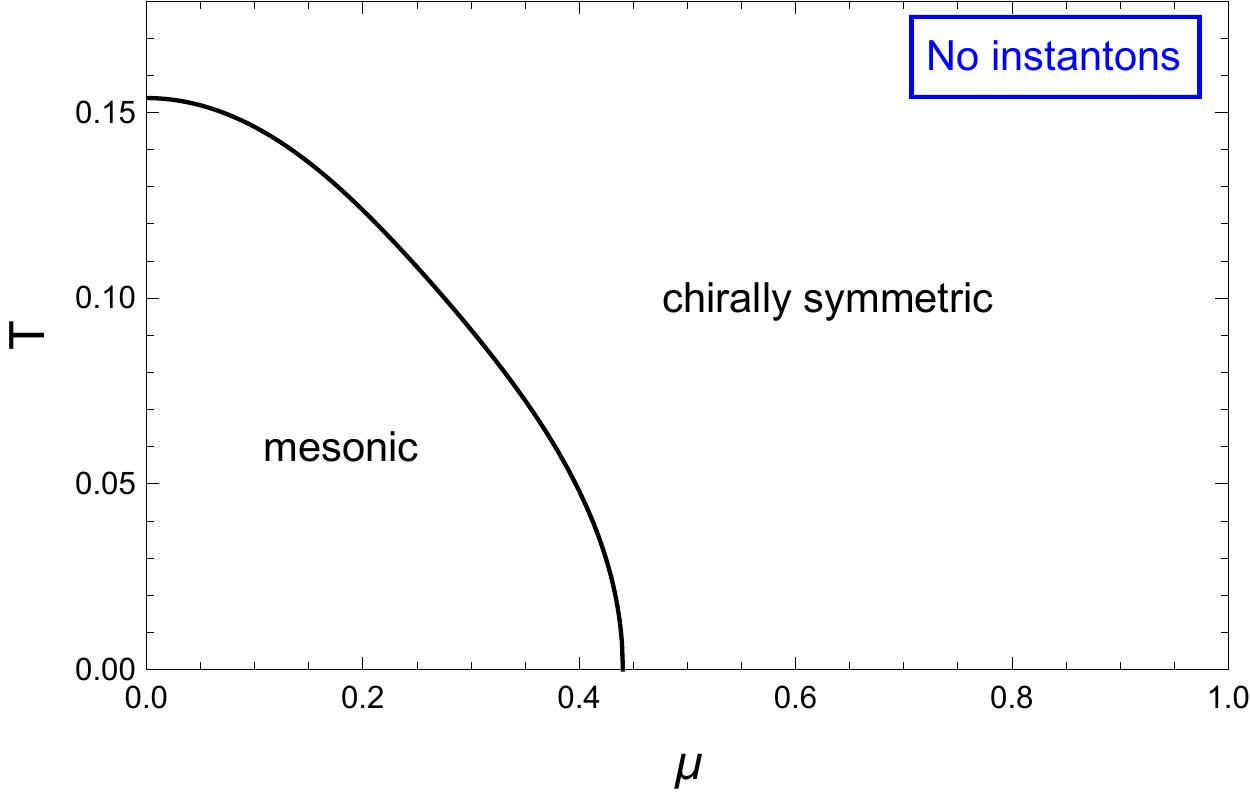}\includegraphics[width=.5\textwidth]{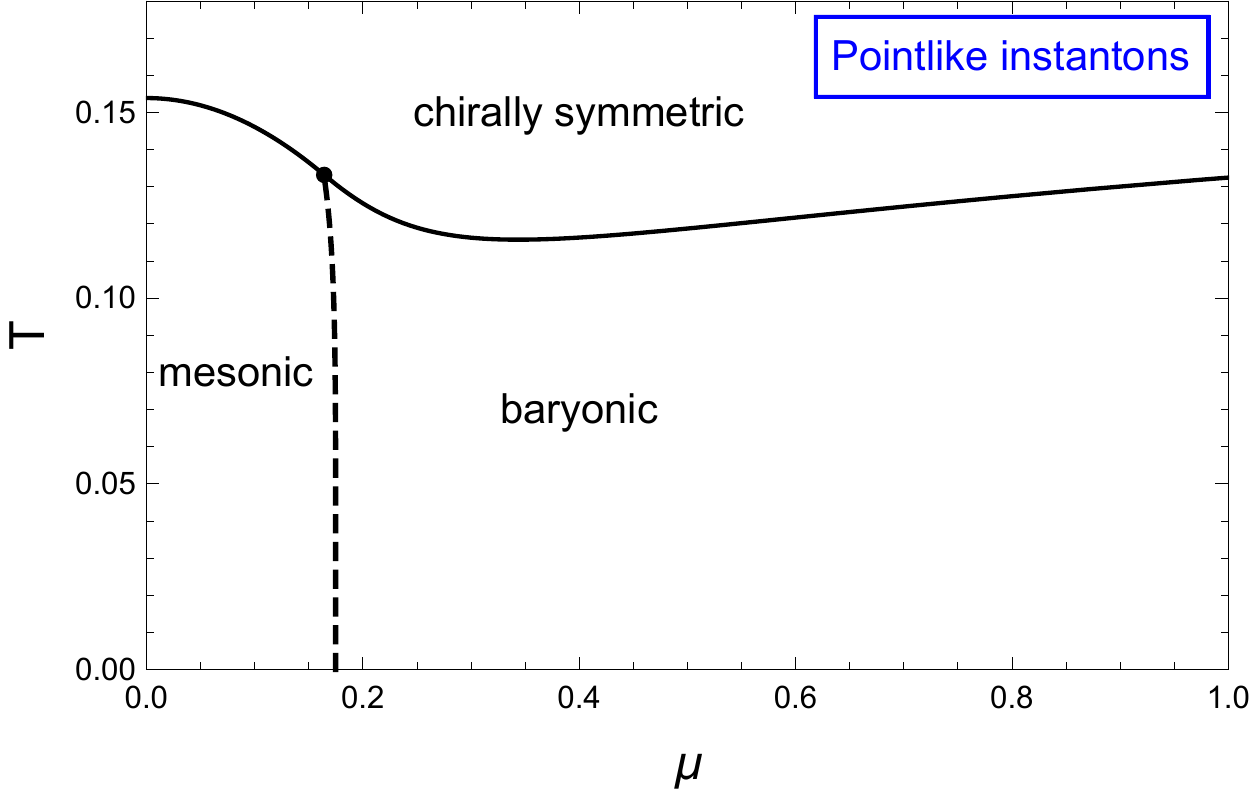}}
\caption{$T$-$\mu$ phase diagrams in the decompactified limit of the Sakai-Sugimoto model, based on previous studies. Solid (dashed) lines are first-order (second-order) 
phase transitions. Left panel: without baryons there is a chirally broken phase
with zero baryon number ("mesonic") and a chirally symmetric quark matter phase \cite{Horigome:2006xu}. Right panel: baryonic matter in the pointlike approximation shows a second-order baryon onset, and chiral 
symmetry is, at low temperatures, not restored at any $\mu$ \cite{Bergman:2007wp,Preis:2011sp}. Here and in the following phase diagrams, we use a dimensionless temperature (the physical temperature is $T/L$) and a dimensionless 
chemical potential (the physical chemical potential is $\mu \lambda /(4\pi M_{\rm KK} L^2)$). 
The present work goes beyond the approximations used for these phase diagrams by allowing for finite-size -- and deformed -- instantons, and includes the possibility of more than one instanton layer.      
}
\label{fig:intro}     
\end{figure}

We employ various approximations for the baryonic phase. 
Firstly, our approach is based on the flat-space instanton 
solution for the non-abelian SU(2) gauge fields on the connected D8- and $\overline{\rm D8}$-branes, and we introduce a many-instanton system by adding up the squared field strengths of the single instantons. 
Moreover, we average in position space over these configurations before we solve the equations of motion for the remaining fields. 
For a single baryon, more sophisticated, but purely numerical solutions are known, and known to be more realistic when compared to real-world nucleons \cite{Cherman:2009gb,Bolognesi:2013nja,Rozali:2013fna}. 
We give up on some degree of sophistication because we have to deal with additional complications, being interested in nuclear matter, not a single nucleon, and, since we work in the deconfined geometry, having to determine the embedding function of the connected flavor branes dynamically. The study of nuclear matter in the Sakai-Sugimoto 
was initiated by approximating the instantons by delta functions \cite{Bergman:2007wp}, see phase diagram in Fig.\ \ref{fig:intro}, and the instanton approach has since then been further developed, including finite-size effects \cite{Ghoroku:2012am,Li:2015uea}. As an alternative approach, not further discussed here, one can use a homogeneous ansatz for the gauge fields which does not rely on any instanton 
solution \cite{Rozali:2007rx,Li:2015uea,Elliot-Ripley:2016uwb}.

\section{Setup of the calculation}
\label{sec:setup}

Our goal is to determine the phase structure of the model 
at nonzero $T$ and $\mu$, with an emphasis on the low-temperature regime, having in mind astrophysical applications. To this end, we focus on the three phases shown schematically in Fig.\ \ref{fig:3cylinders}, and 
compare their free energies. The mesonic and quark matter phases are simply taken from the literature (see for instance appendix B of Ref.\ \cite{Li:2015uea} for a brief review of these phases, in the same notation as used below). The most difficult phase, where various approximations are necessary, is the baryonic phase. We shall discuss this phase in the following, explaining the setup and the calculation very briefly, all details can be found in Ref.\ \cite{Preis:2016fsp}, 
on which the results of these proceedings are based. 

The Lagrangian we start from consists of a Dirac-Born-Infeld part and a Chern-Simons part, 
\be \label{Lag}
{\cal L} = u^{5/2} \sqrt{(1+u^3f_T x_4'^2-\hat{a}_0'^2+g_1)(1+g_2)} - n_I\hat{a}_0 q \, ,
\ee
where $u$ is the holographic coordinate, prime denotes derivative with respect to $u$, $n_I$ is the instanton density, $f_T = 1-u_T^3/u^3$ with $u_T = \left(\frac{4\pi T}{3 M_{\rm KK}}\right)^2$, $\hat{a}_0$ is the 
abelian gauge field with boundary condition $\hat{a}_0(\infty) = \mu$, and $x_4(u)$ is the embedding function of the flavor branes with boundary conditions $x_4(u_c)=0$, $x_4(\infty)=L M_{\rm KK}/2$. 
We work exclusively with dimensionless quantities, for the corresponding dimensionful forms see Ref.\ \cite{Preis:2016fsp}. 
The functions $g_1$, $g_2$, and $q$ arise from the non-abelian field strengths, $F_{iu}^2$, $F_{ij}^2$, and $\epsilon_{ijk} F_{ij}F_{ku}$, and they are based on the flat-space instanton approximation, averaged over position space. Explicitly, we have
\be\label{g1g2sim}
g_1(u) \equiv  \frac{f_T n_I}{3\gamma}\frac{\partial z}{\partial u}q(u) \, , \qquad g_2(u) \equiv  \frac{\gamma n_I}{3u^3} \frac{\partial u}{\partial z} q(u)  \, ,
\ee
where the variables $z$ and $u$ are related by $u = (u_c^3+u_c z^2)^{1/3}$, and 
\be
q(u) \equiv 2\frac{\partial z}{\partial u} D(z) \, , \qquad \int_{u_c}^\infty du\, q(u) = 1 \, .
\ee
The parameter $\gamma$ accounts for the anisotropy of the instantons. In flat space, corresponding to the \mbox{$\lambda=\infty$} limit, they are SO(4) symmetric in $(\vec{x},z)$ 
space, with $\vec{x}=(x_1,x_2,x_3)$. The deformation parameter $\gamma$ is a simple way -- keeping the functional
form of the flat-space solution -- to allow for oblate and prolate instantons. It is known from the full numerical solution that, upon going beyond the $\lambda=\infty$ limit, such a deformation does set in. In our simple parameterization, the instanton becomes elongated along the holographic direction for large $\gamma$ and along the spatial direction $x\equiv |\vec{x}|$ for small $\gamma$ (the instantons remain SO(3) symmetric in 
position space). The function $D(z)$ can be understood as the instanton profile on the flavor branes. It includes $N_z$ many instanton layers that are allowed to spread along the $z$ direction,
\be \label{Dz}
D(z) \equiv \int d^3 x \,D(x,z) \, , \qquad D(x,z) \equiv \frac{6}{\pi^2\gamma N_z}\sum_{n=0}^{N_z-1}\frac{(\rho/\gamma)^4}{[x^2+(z-z_n)^2/\gamma^2+(\rho/\gamma)^2]^4} \, ,
\ee
where $\rho$ is the instanton width (giving, in combination with the deformation parameter $\gamma$, two different widths in holographic and spatial directions). We speak of instanton {\it layers} because, in general, they 
also have a distribution in the spatial direction. The form of this distribution (for example a certain lattice structure) is irrelevant in our approximation because we neglect their interaction in position space and spatially average 
over the field strengths squared. The locations of the instanton layers in the bulk is denoted by $z_n$,
and we assume them to be distributed equidistantly and symmetrically around $z=0$, 
\be \label{zn}
z_n= \left(1-\frac{2n}{N_z-1}\right) z_0 \, ,
\ee
where the parameter $z_0$ (as well as the number $N_z$) has to be determined dynamically. 

The free energy of the baryonic phase 
\be
\Omega = \int_{u_c}^\infty du\, {\cal L} 
\ee
is computed by first solving the equations of motion for $\hat{a}_0(u)$ and $x_4(u)$. We can solve them algebraically for  $\hat{a}_0'(u)$ and $x_4'(u)$,
\bea
\hat{a}_0'^2 &=&  \frac{(n_IQ)^2}{u^5}\frac{1+g_1}{1+g_2-\frac{k^2}{u^8f_T}+\frac{(n_IQ)^2}{u^5}} \, , \qquad x_4'^2 =  \frac{k^2}{u^{11}f_T^2}\frac{1+g_1}{1+g_2-\frac{k^2}{u^8f_T}+\frac{(n_IQ)^2}{u^5}} \, , \label{dx4}
\eea
where $k$ is an integration constant, and 
\be
Q(u) \equiv \int_{u_c}^u dv\,q(v) 
\, .
\ee
The solutions are inserted back into $\Omega$, and it remains to minimize the resulting expression with respect to the free parameters at given $T$ and $\mu$,
\be
0 = \frac{\partial \Omega}{\partial k} =\frac{\partial \Omega}{\partial n_I} =\frac{\partial \Omega}{\partial u_c} =\frac{\partial \Omega}{\partial z_0} =\frac{\partial \Omega}{\partial \rho} =\frac{\partial \Omega}{\partial \gamma}  \,,  
\ee
plus finding the integer $N_z$ that simultaneously minimizes $\Omega$. 

If the baryonic phase is treated exactly as just outlined, the zero-temperature results are as follows: there is a second-order baryon onset (= $n_I$ continuous), 
just like for the pointlike approximation (in fact, the instantons do become 
pointlike, $\rho\to 0$ for $n_I\to 0$), and there is no chiral restoration at large $\mu$, which, again, was already seen in the pointlike approximation. Both features are unphysical in the sense that QCD behaves differently. 
For any sensible equation of state, which accounts for the basic physical 
properties of real-world baryons and which is capable to allow for hybrid stars, i.e., neutron stars with a quark matter core, a first-order baryon onset and chiral restoration at large $\mu$ are necessary.
It is not obvious whether the full solution of the Sakai-Sugimoto model does behave like QCD with regard to the baryon onset and chiral restoration. Therefore, a given approximation should not necessarily be judged
by these physical properties. More important from a theoretical point of view, the solution only has a single instanton layer, we have not found any solution for $N_z>1$. 
The physical meaning of the instanton layers in the bulk is not obvious, but from this observation 
we {\it can} conclude that our approximation is too simplistic; at least it is strongly suggested from other, complementary, approaches that the instanton distribution should spread out in the holographic direction 
at large densities \cite{Rozali:2007rx,Elliot-Ripley:2016uwb,Kaplunovsky:2012gb}.  

It turns out that the instantons do spread out dynamically if we improve our approximation by imposing the following constraints on the instanton width and deformation,
\be\label{rhogamma}
\rho = \rho_0 u_c \, , \qquad \gamma = \frac{3}{2}\gamma_0u_c^{3/2} \, ,
\ee
with $\rho_0$ and $\gamma_0$ now externally fixed, increasing the number of parameters of the model to five (besides $\lambda$, $M_{\rm KK}$ and $L$), and $u_c$ dynamically determined, as before. This somewhat 
phenomenological approach accounts for instance for the fact that the instantons always have a nonzero width, as suggested from rigorous single-baryon results at finite $\lambda$. The particular scaling with $u_c$ is motivated by the 
specific form of the stationarity equations, allowing for a complete elimination of $u_c$ from all equations but one. In the rest of these proceedings, we present the results from this approach. 

\section{Results and conclusions}

\begin{figure}[t]
\centering
\hbox{\includegraphics[width=.32\textwidth]{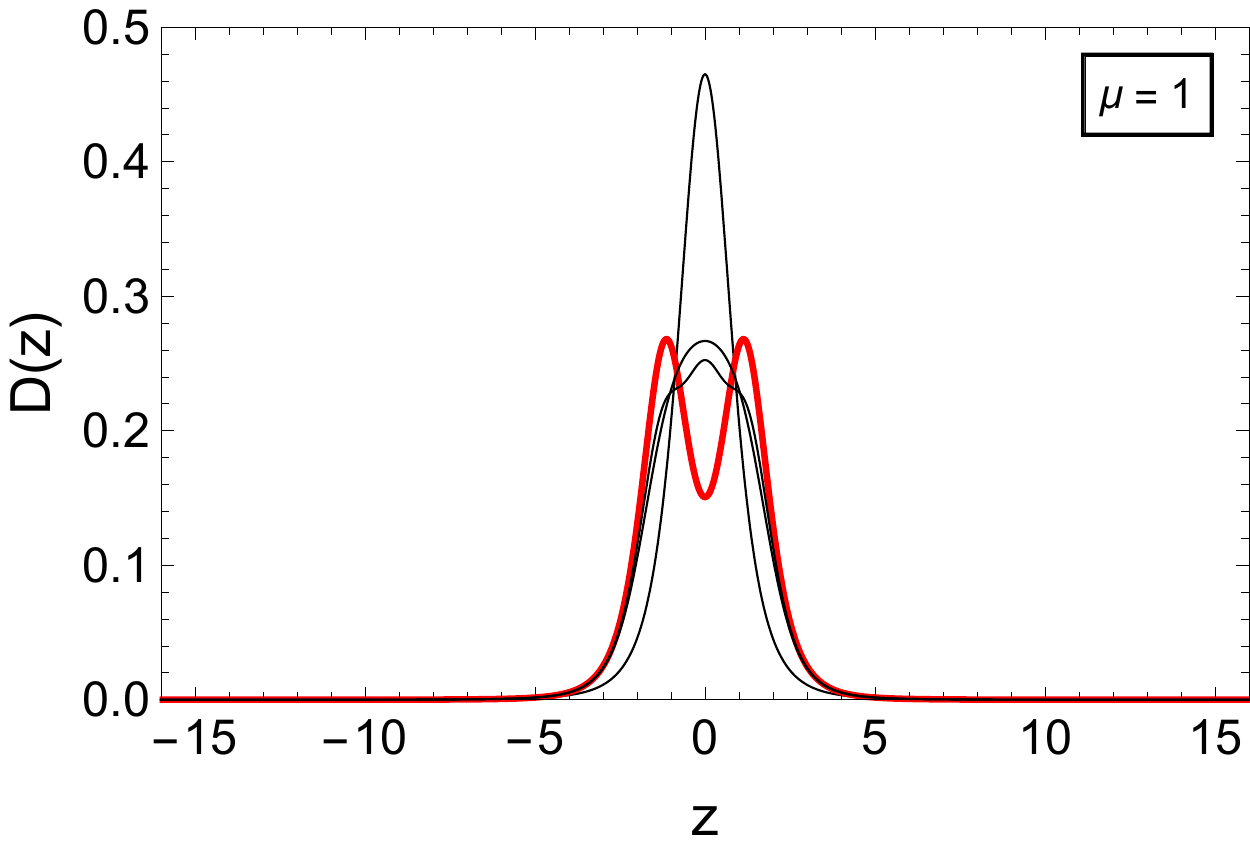}\includegraphics[width=.33\textwidth]{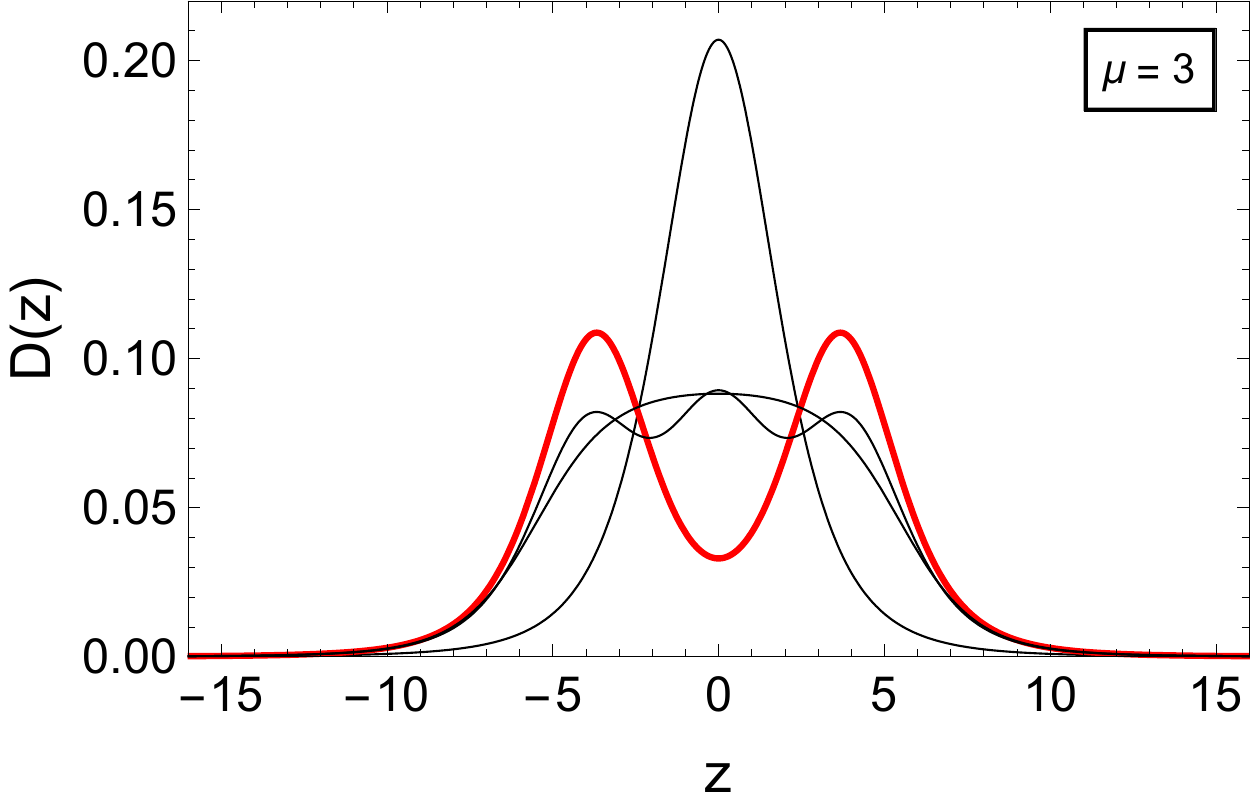}\includegraphics[width=.33\textwidth]{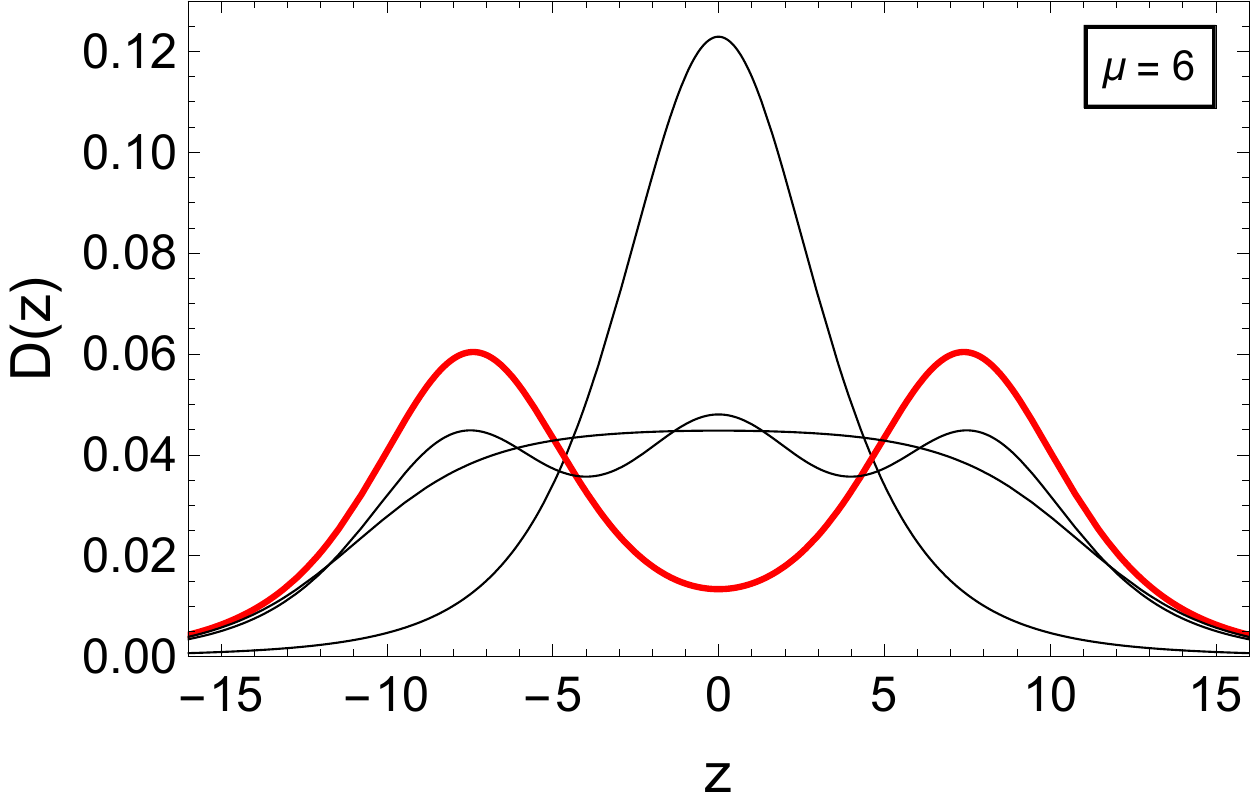}}

\vspace{0.5cm}
\hbox{\includegraphics[width=.33\textwidth]{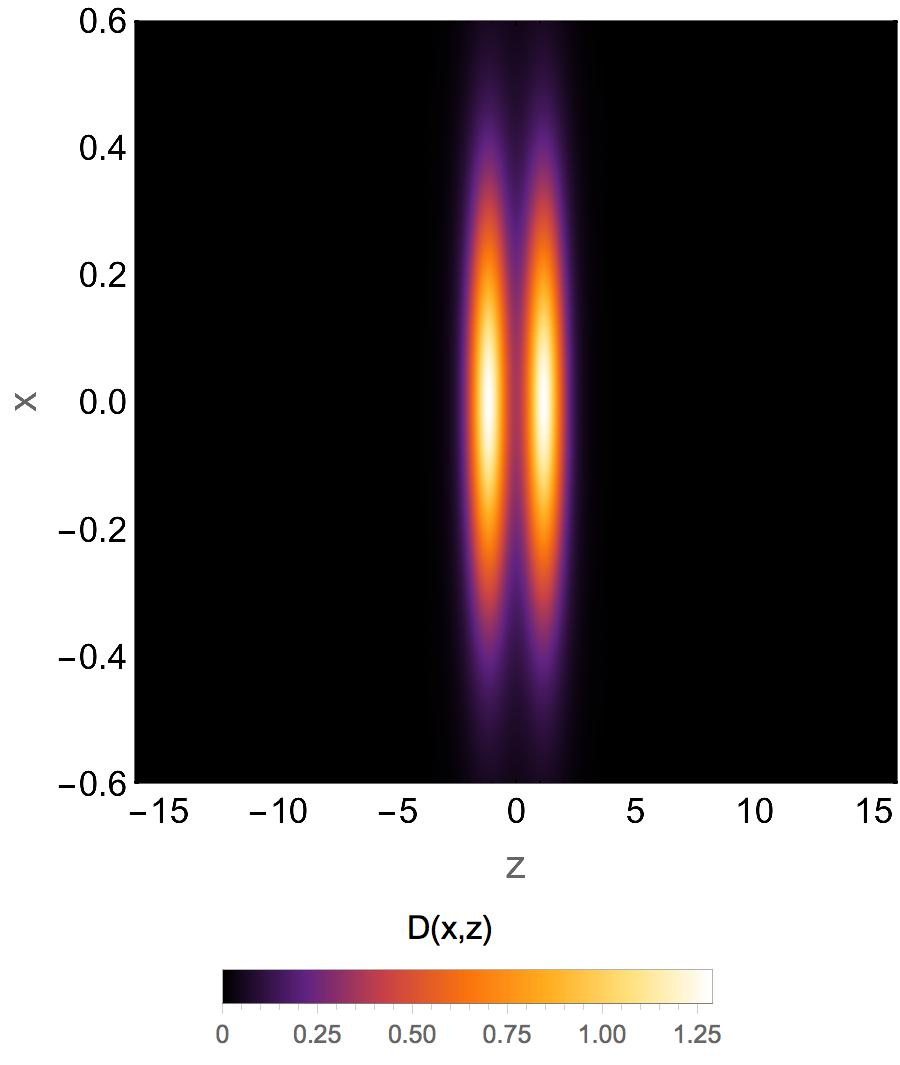}\includegraphics[width=.33\textwidth]{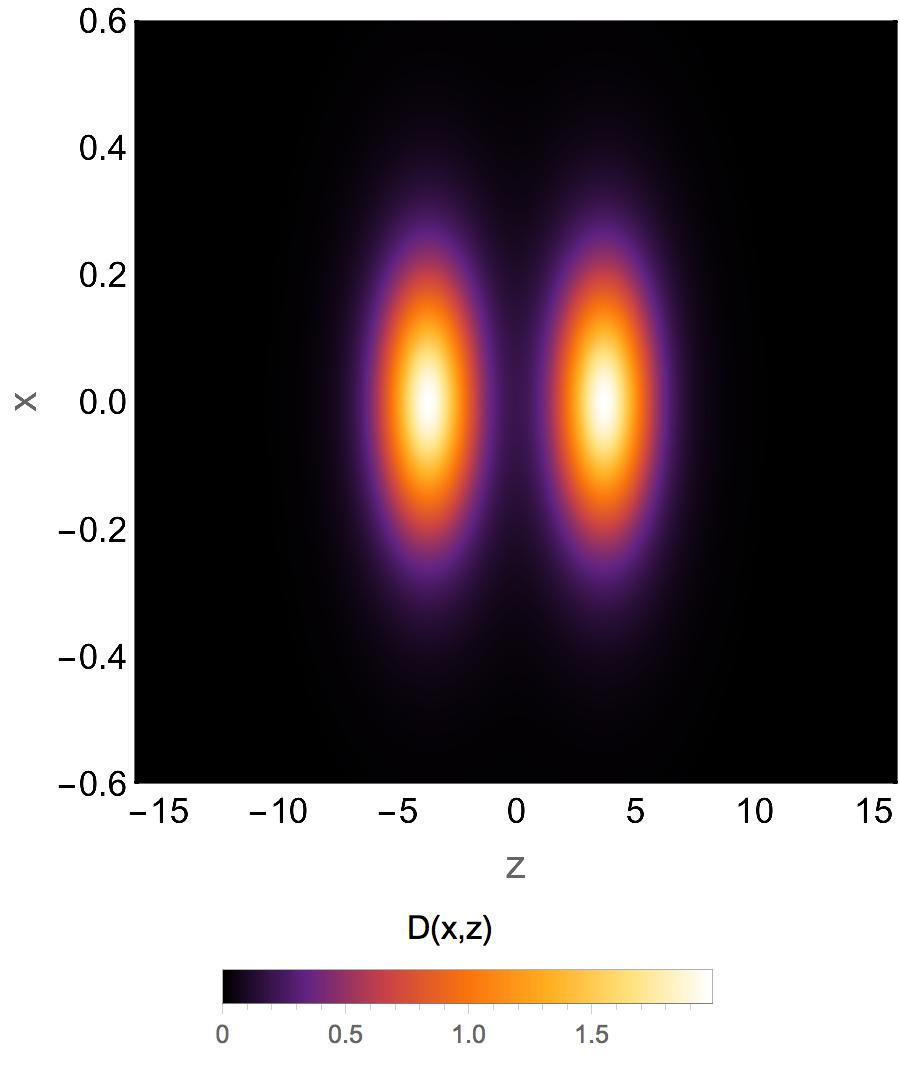}\includegraphics[width=.33\textwidth]{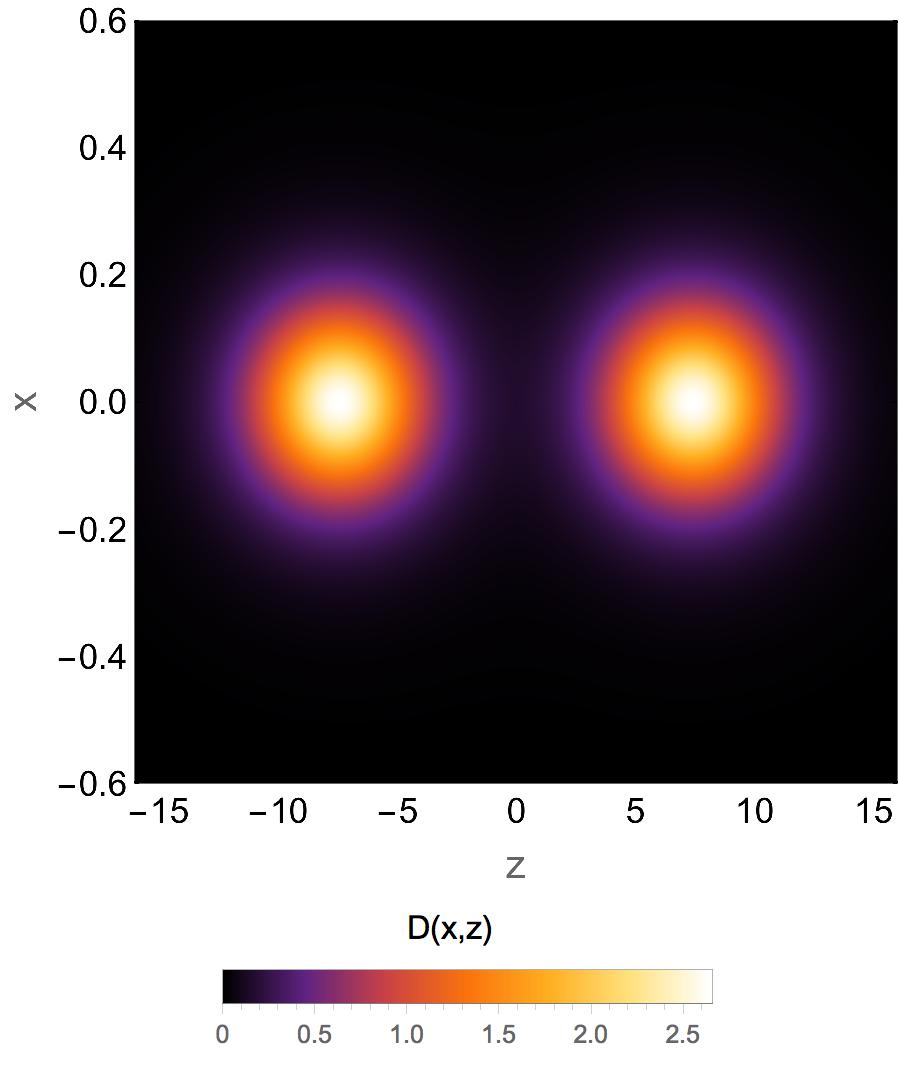}}
\caption{Instanton profiles in the holographic direction $z$ (upper row) and in the $(x,z)$ space (lower row) for three different chemical potentials and $T=0$, with parameters $\rho_0=2.5$, $\gamma_0=4$. 
The thick (red) curves in the upper row
show the energetically preferred two-layer solution (at lower values of $\mu$, only one layer is favored, see right panel of Fig.\ \ref{fig:phases}). 
For comparison, we also show the profiles for $N_z=1,3,\infty$. The lower row shows the two-layer solution for the same three chemical potentials. }
\label{fig:bumps}
\end{figure}

Firstly, we discuss the structure of the solutions in the bulk, before we get to the phenomenologically more relevant properties of the solution. For specific values of $\rho_0$ and $\gamma_0$, we plot the instanton profiles 
given in Eq.\ (\ref{Dz}) for three different chemical potentials, see Fig.\ \ref{fig:bumps}.
 The figure shows the two-layer structure, with the distance between the two layers growing with increasing density. Additionally, the lower row of the figure shows the change in deformation: 
the instanton width in the spatial (holographic) direction shrinks (grows) with increasing density. For all values of $\rho_0$ and $\gamma_0$ (which we have checked) and all densities, $N_z=2$ is the maximum number of 
energetically preferred instanton layers. 
We cannot exclude that this is an artifact of our approximation, for instance by restricting ourselves to equidistant layers in $z$. However, the same result was found in a completely different approximation 
\cite{Elliot-Ripley:2016uwb}.

\begin{figure}[t]
\centering
\hbox{\includegraphics[width=.5\textwidth]{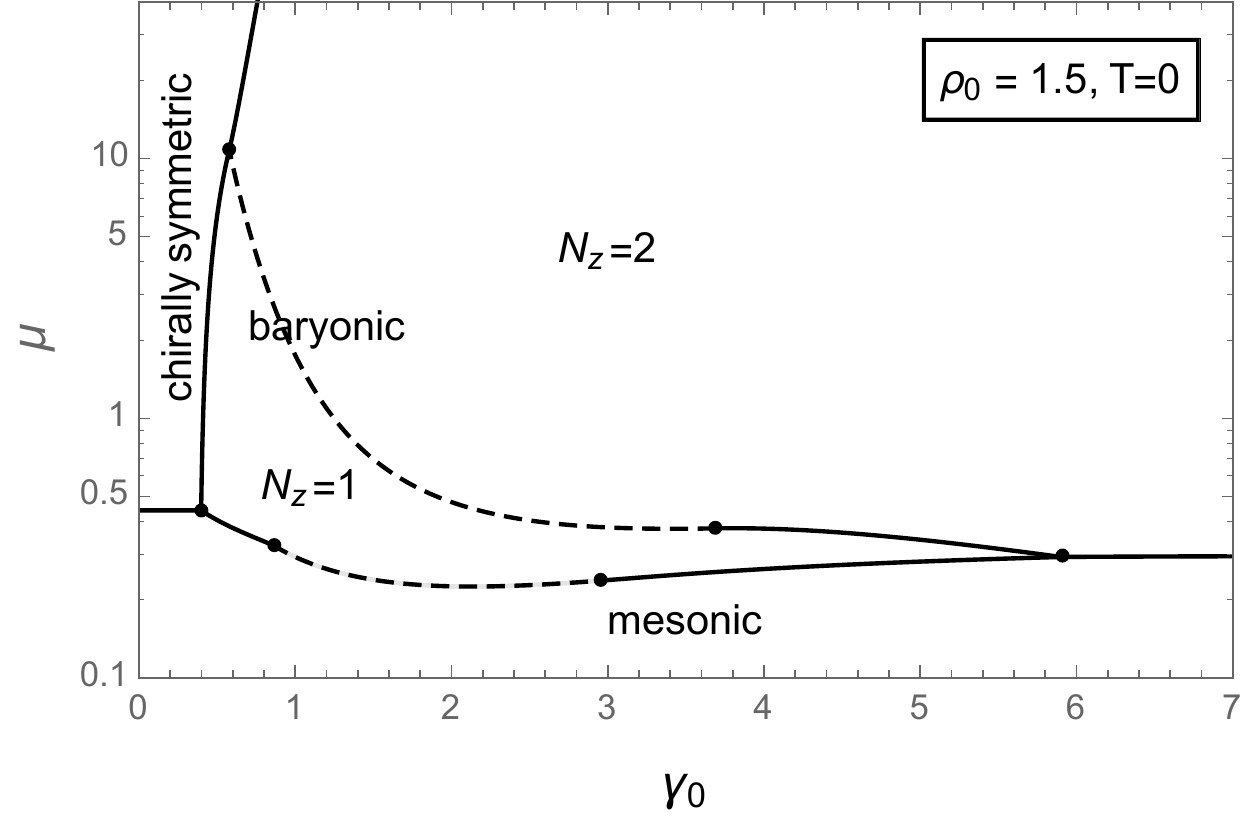}\includegraphics[width=.5\textwidth]{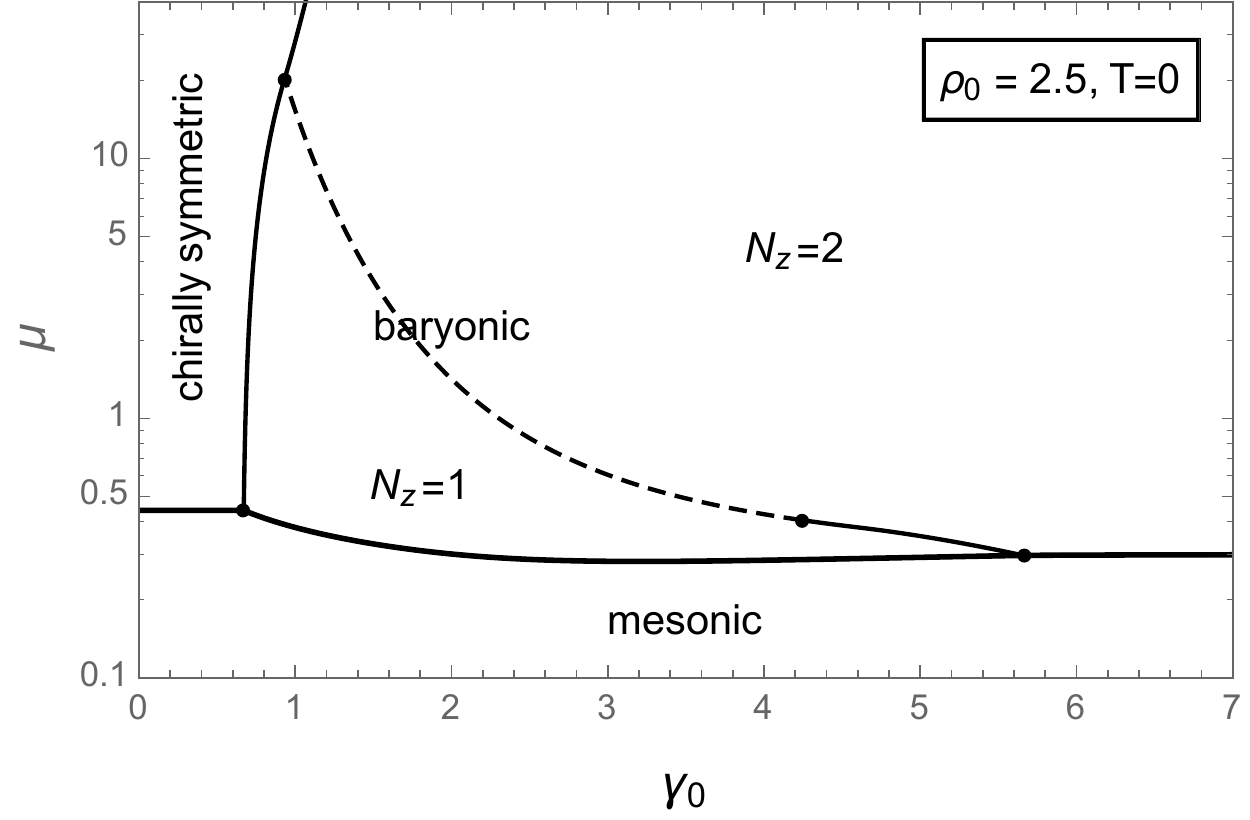}}
\caption{Zero-temperature phases for two slices through the phase space spanned by the parameters $\rho_0$, $\gamma_0$ (which set constraints on the instanton width and deformation) and the chemical potential
$\mu$. Solid (dashed) lines are first-order (second-order) phase transitions. The number of instanton layers $N_z$ in the baryonic phase is either one or two, and the transition between them can either be 
continuous or discontinuous.} 
\label{fig:phases} 
\end{figure}

While we take the results of Fig.\ \ref{fig:bumps} as an indication of the improvement of our approximation through the constraints on $\rho$ and $\gamma$, let us now see whether we have also improved
the phenomenological properties of our baryonic matter. We start with the zero-temperature results, for two given values 
of $\rho_0$, varying the second parameter $\gamma_0$ arbitrarily. The results are shown in Fig.\ \ref{fig:phases}. We see that there {\it is} a parameter range in which the baryon onset is first order, i.e., there is a binding energy 
for nuclear matter. We also see that there is a (very small) region where, upon increasing $\mu$ at fixed $\rho_0$ and $\gamma_0$, the baryon onset is followed by a chiral phase transition to quark matter. In principle, one might proceed and match the 5 parameters of the model to physical data, for instance to saturation properties of nuclear matter. As a result, our model would predict the location of the chiral phase transition 
to quark matter (or the absence thereof). This possibility is remarkable, we are not aware of any other model with a comparably small number of parameters, where low-density and high-density properties can be coupled in such a rigid way (usually, different models for the different density regimes have to be employed). Of course, our model has to be taken with a lot of care because of the above mentioned approximations, some of which are very crude. 
In particular, it would be important to include the interaction of the instantons in position space, not only in the holographic direction, where we already have seen the repulsion of instanton layers. 
Such an interaction would presumably give rise to an increased energy cost for the baryons to overlap, and thus possibly to a transition to quark matter at a lower chemical potential (or to the existence of a 
such a transition in parameter regions where it does not exist in the present treatment). We are currently working on such an extension \cite{preparation}, and therefore defer a full-fledged evaluation of the model, including a matching of the parameters, to future work. 

\begin{figure}[t]
\centering
\includegraphics[width=.5\textwidth]{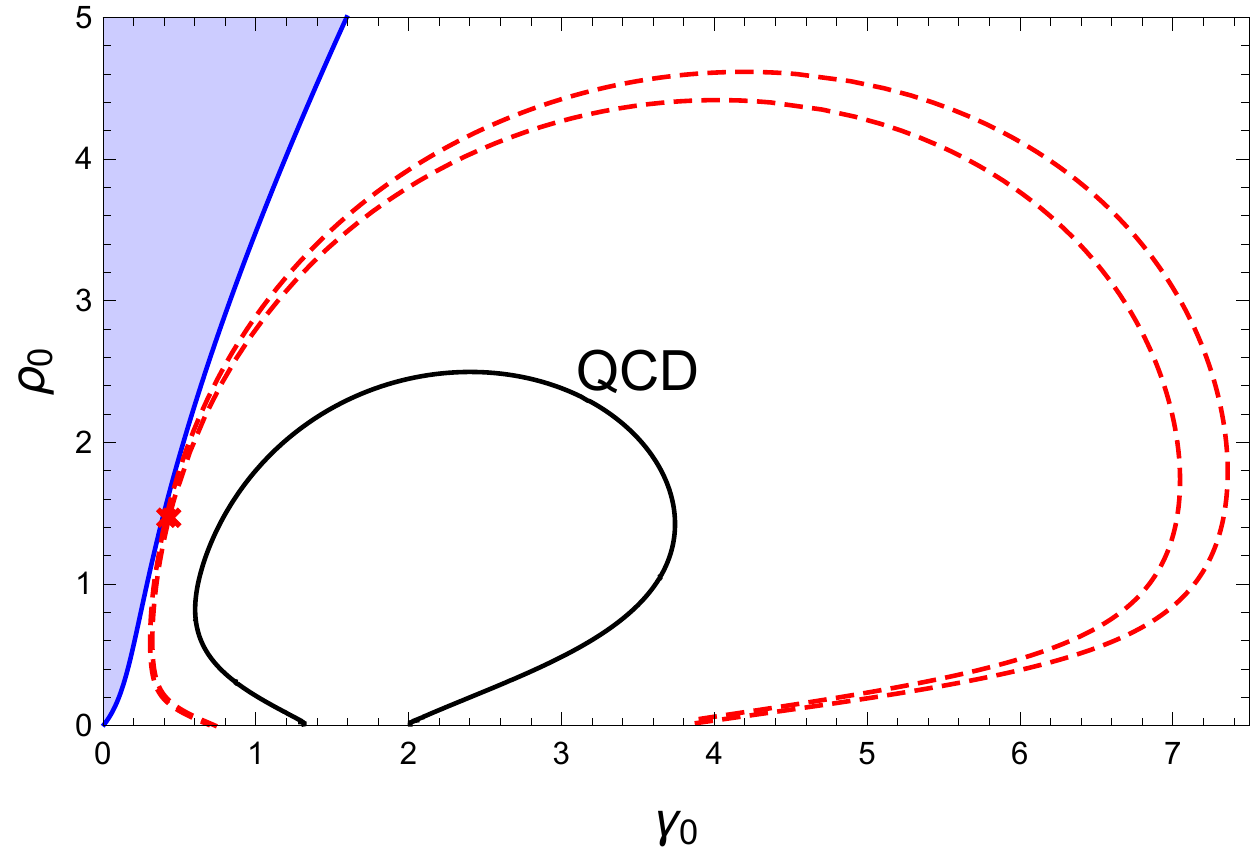}
\caption{Curve in the $\rho_0$-$\gamma_0$ parameter space on which the zero-temperature binding energy of symmetric nuclear matter is as small as in QCD, $\mu_0/M_0 = 98\%$
[(black) solid line].  
In the shaded region there is no baryonic matter for any chemical potential, i.e., the curve that bounds this region is the location of the triple point in the phase diagrams of Fig.\ \ref{fig:phases}, where baryonic, mesonic, and chirally symmetric phases meet. For comparison, we show two curves [(red) dashed] with unrealistically large 
relative binding energies, $\mu_0/M_0 = 73\%$ and 75\%. The (barely distinguishable) crosses on these two curves
mark the points that are chosen for the phase diagrams in Fig.\ \ref{fig:Tmuphases}. While they show a $T=0$ chiral phase transition at high density, the 
points on the "QCD curve" have either no $T=0$ chiral phase transition at all or only at extremely large $\mu$ (possibly a consequence of our 
simple approximation that appears to treat baryonic matter too favorable). } 
\label{fig:r0g0} 
\end{figure}

Here we continue with one aspect of a more quantitative evaluation, namely the strength of the first-order baryon onset, i.e., the size of the binding energy. We know that for symmetric 
nuclear matter this binding energy is about 16 MeV, much less than the vacuum mass of the baryon $M_0 = 939\, {\rm  MeV}$. In other words, the chemical potential for the baryon onset $\mu_0$ is only slightly smaller than the vacuum mass. Does our model allow for binding energies that are small in this sense, i.e., $\mu_0/M_0 \simeq 98\%$? Fig.\ \ref{fig:phases} suggests that it does because it contains a second-order
phase transition line (where the binding energy is zero) which terminates in two critical points and continues as first-order transition lines. 
This implies that the jump in the baryon density becomes arbitrarily small close to the two 
critical points, and consequently we should be able to find parameter values for $\rho_0$ and $\gamma_0$ for any given small binding energy. The dimensionless ratio $\mu_0/M_0$ is interesting also because all other parameters cancel, i.e., the condition of a given binding energy translates into a condition for $\rho_0$ and $\gamma_0$ alone. In our approach, we can calculate the vacuum mass of a baryon most easily by computing the  chemical potential for vanishing density, $M_0 = N_c\mu(n_I\to 0)$ (the factor $N_c$ arises because $\mu$ is the {\it quark} chemical potential). We can thus, together with the calculation of the critical chemical potential 
for the baryon onset $N_c\mu_0$,
determine a curve in the $\rho_0$-$\gamma_0$ parameter space on which the QCD binding energy is reproduced (more precisely, the {\it relative} binding energy, reproducing the correct {\it absolute} values 
of $\mu_0$ and $M_0$ would require fitting one more parameter). The result is shown in Fig.\ \ref{fig:r0g0}.

\begin{figure}[h]
\centering
\hbox{\includegraphics[width=.5\textwidth]{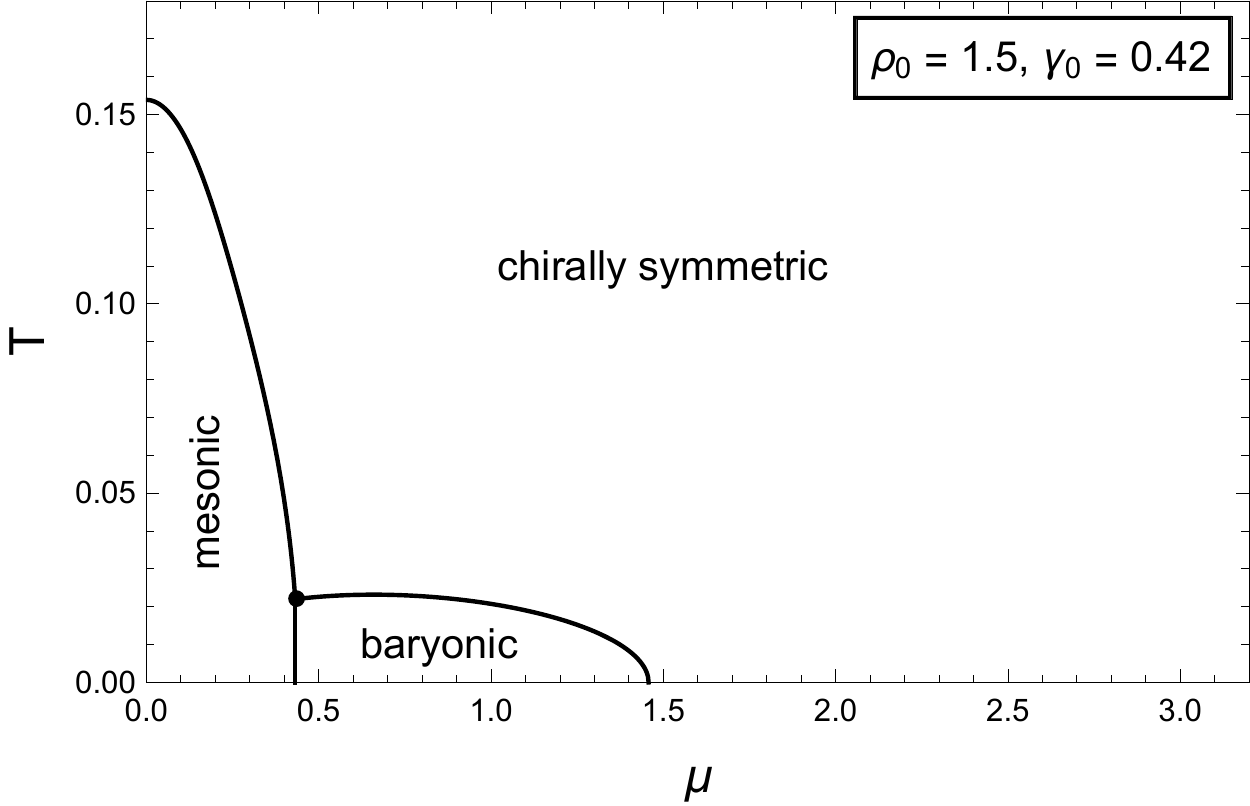}\includegraphics[width=.5\textwidth]{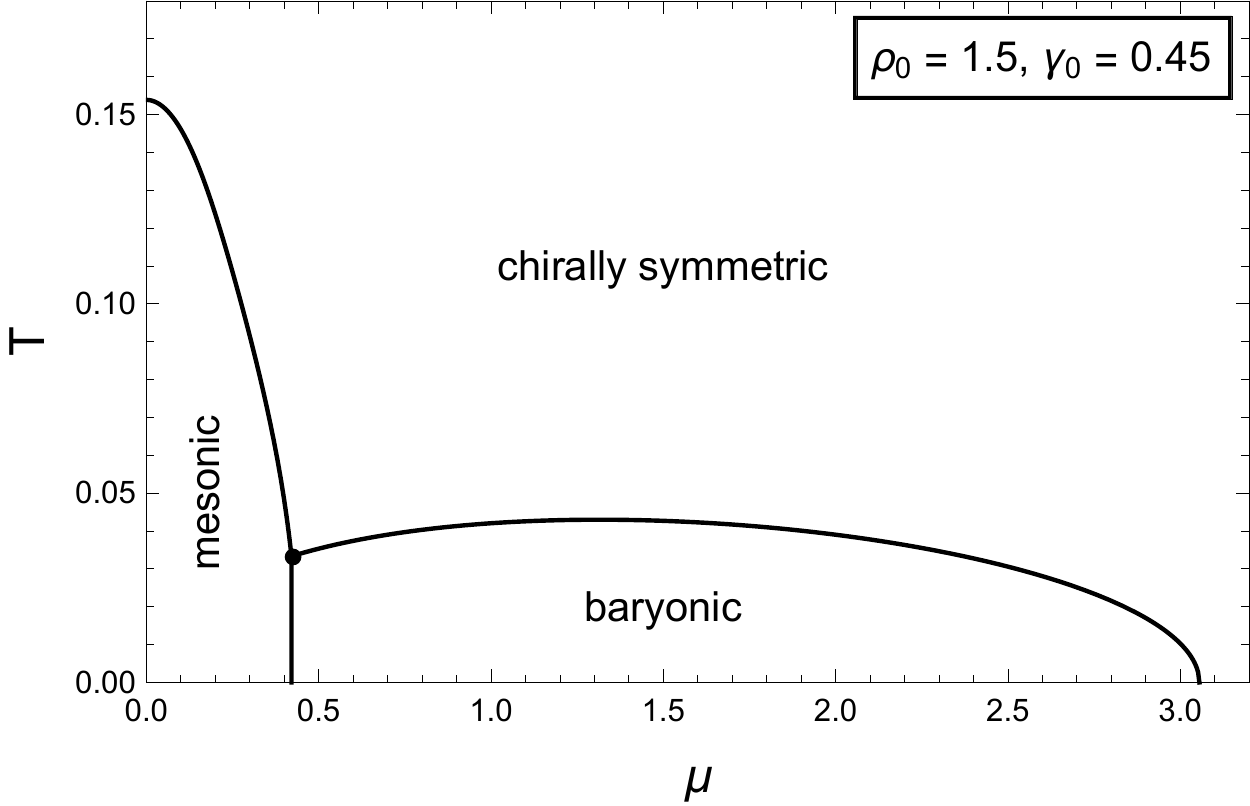}}
\caption{Phase diagram in the plane of temperature and chemical potential for a fixed $\rho_0$ and two values for $\gamma_0$.}
\label{fig:Tmuphases}     
\end{figure}

Finally, we explore the full phase structure in the $T$-$\mu$ plane in Fig.\ \ref{fig:Tmuphases}. We have chosen the value of $\rho_0$ from the left panel of Fig.\ \ref{fig:phases}, and two values of $\gamma_0$
for which there is a zero-temperature phase transition to quark matter. As already obvious in Fig.\ \ref{fig:phases}, this requires some fine-tuning of the parameters, and the $T$-$\mu$ phase diagrams confirm that the
chiral phase transition is very sensitive to the choice of $\gamma_0$.  Notice that the zero-temperature baryon onset
remains a transition from the mesonic phase to the baryonic phase at non-zero temperatures, i.e., the baryon number is always zero for chemical potentials below that transition. In QCD we expect a liquid/gas phase transition, 
i.e., except for the exact $T=0$ limit, there is always a nonzero baryon density (albeit exponentially suppressed at small temperatures). Although our main focus has been on the zero-temperature properties, a better understanding of this $T>0$ behavior is another interesting problem for future studies.

{\it Acknowledgments.} 
We acknowledge support from the {\mbox NewCompStar} network, COST Action MP1304. 
A.S.\ is supported by the Science \& Technology Facilities Council (STFC) in the form of an Ernest Rutherford Fellowship.

\bibliography{references}

\end{document}